\documentclass[12 pt]{article}

\usepackage{enumerate,amsthm,amsmath,amssymb,color,graphicx,bbm,epsfig,bbm}

\def\1{\mathbf{1}}

\title{Security of Trusted Repeater Quantum Key Distribution Networks}

\author{Louis Salvail$^1$,  Momtchil Peev$^2$, Eleni Diamanti$^{3,4}$, Romain All\'eaume$^4$,  \\
Norbert L\"utkenhaus$^{5,6}$, Thomas L\"anger$^2$ \\ \\
$^1$ {\em Universit\'e de Montr\'eal, Montr\'eal, Canada} \\
$^2$ {\em Austrian Research Centers GmbH - ARC, Vienna, Austria}\\
$^3$ {\em Laboratoire Charles Fabry de l'Institut d'Optique, Palaiseau, France}\\
$^4$ {\em Telecom ParisTech \& LTCI - CNRS, Paris, France} \\
$^5$ {\em University Erlangen-Nuremberg, Erlangen, Germany} \\
$^6$ {\em Institute for Quantum Computing, University of Waterloo, Waterloo, Canada}  \\
}

\begin{document}

\maketitle

\abstract{A Quantum Key Distribution (QKD) network is an infrastructure capable of performing
long-distance and high-rate secret key agreement with information-theoretic security. In this paper
we study security properties of QKD networks based on trusted repeater nodes. Such networks can
already be deployed, based on current technology. We present an example of a trusted repeater
QKD network, developed within the SECOQC project.
 The main focus is put on the study of secure key agreement over a trusted repeater QKD network, when some
 nodes are corrupted. We propose an original method, able to ensure the authenticity and privacy of
 the generated secret keys.}

\section{Introduction}
\label{sec:intro}

Quantum Key Distribution (QKD), often called in a more general context Quantum Cryptography, is a
technology that uses the properties of quantum mechanical systems in combination with information
theory to achieve unconditionally secure distribution of secret keys. In the last years, the field
has rapidly evolved in terms of both  theoretical foundations and experimental implementations,
with impressive results~\cite{gisin:rmp02,dusek:pino06,scarani:rmp08}.

 The use of QKD has been, until now, mostly limited to point-to-point communication scenarios: the
 goal being to allow two remote parties linked by a quantum channel and an authentic classical
 channel to share a common random binary string - a key - that remains unknown to a potential
 eavesdropper, and to achieve in practice the longest possible communication distance and the
 highest possible key generation rate. Despite the progress in this direction, the performance of
 stand-alone point-to-point QKD links will however remain intrinsically limited in terms of
 achievable distance and rate. Building QKD networks based on an ensemble of QKD links and
 intermediate nodes, could lift these limitations.  The purpose of this paper is to discuss the
 security aspects of  QKD networks whose deployment is feasible with current technology: trusted
 repeater QKD networks. The principle of such networks consists in using trusted repeater nodes as
 classical relays between QKD links. Indeed, provided that {\em some} level of trust can be granted to
 the network nodes, such networks can guarantee  unconditionally secure key exchange between
 multiple users over potentially unlimited distances.

The material is organized  as follows. Section~\ref{sec:KEP_QKD} defines the setting of this work:
key agreement based on Quantum Key Distribution. It introduces the cryptographic framework of
Quantum Key Distribution, focusing on its most striking cryptographic feature: the ability to
establish secret keys with information-theoretic security. Section \ref{sec:QKDnetworks} then
describes the different possible types of QKD networks and presents an example of a trusted
repeater QKD network: the {\sc Secoqc} QKD network.  Section~\ref{sec:section4} is then devoted to
the full analysis of secure key agreement in a trusted repeater QKD network in the case when some
nodes may be arbitrarily malicious (or corrupt). We propose a method allowing the communicating
parties to ensure the authenticity of a generated secret key without compromising its privacy. We
also discuss practical issues and provide a security analysis for this method. Finally, in
Section~\ref{sec:perspectives}, we summarize our results and discuss possible modifications in the
model assumptions.

\section{The Key Establishment Problem and Quantum Key Distribution}
\label{sec:KEP_QKD}
In this work, we regard QKD as a cryptographic primitive, that is as a
low-level, universal cryptographic algorithm which can be used as a building block for creating
highly complex, dedicated secure communication applications. In this sense, the task of QKD is key
distribution (or to use the proper cryptographic term {\it key establishment}) between two
legitimate parties at two distant locations.

Key Establishment~\cite{Menezes} is a standard security task, which  is solved either by sending
the key from one party to the other over a channel assumed to be secure ({\it key transport}) or by
applying methods allowing the two parties to generate a common secret key out of inputs provided by
both parties ({\it key agreement}). Key establishment methods are based on protocols, including
specific, locally executed, algorithmic steps and public communication. Assumptions on the
intrinsic properties of the communication channels, the power of the adversary, or the resources
available to the legitimate parties  yield a variety of models, which depending on the methods
applied offer different levels of security.

In Section~\ref{sec:ITS} we introduce information-theoretic security - a security level, provided
by QKD,  which is also central to all protocols discussed in this paper.
Section~\ref{sec:ITSmodels+QKD} gives a short overview of models allowing information-theoretic
security followed by a detailed discussion of the crypto-properties of QKD, which are  the corner
stone of the subsequent constructions. Section~\ref{sec:QKDLinks} addresses then performance and
applicability issues of typical realizations of this primitive and argues on the necessity of
designing QKD networks.

\subsection{Information-Theoretically Secure  Key Agreement}
\label{sec:ITS}

 It is beyond the scope of the current paper to address in detail all possible levels of security
of  key establishment models. We will be solely interested in the highest level of security, known
as {\it information-theoretic} (or {\it unconditional}) security. The notion of
information theoretic security (ITS), which is based on probability-theoretic statements,  goes
back to Shannon~\cite{Shan1, Shan2}. This notion was first introduced in the context of a key
agreement process by Wyner~\cite{Wyner}. An exact definition depends on the precise model
assumptions. Here we describe ITS key agreement in general terms (following ~\cite{wolf:phd99}) of
the two underlying essential ingredients Authenticity and Privacy.

Two parties Alice and Bob perform a key establishment process, as a result of which they obtain the
keys $K^A$ and $K^B$ respectively of length $n$. We say that the key agreement is
$\epsilon(n)$-secure if there exists a perfectly random, uniformly distributed key $K$ of length
$n$, for which
        \begin{description}
            \item[i. (Authenticity):] The probability  that 
            $\left((K^A\neq K) \bigvee \,(K^B\neq K)\right) \leq \epsilon'(n)$ OR the key agreement
            process is terminated with notification of failure,
            \item[ii. (Privacy):]The information of the adversary Eve\footnote{Here the information of the adversary is
            symbolically meant in a generic sense. Strictly speaking
            mutual information $I(K:E)$ is defined only in the sense of Shannon entropy, i.e.
            when the the knowledge of the adversary can be characterized by a probability distribution. See Section~\ref{sec:ITSmodels+QKD} for an adversary holding quantum information.}
            is bounded by $I(K:E) \leq \epsilon''(n),$
        \end{description}

\noindent whereby $\epsilon(n)=\epsilon'(n)+\epsilon''(n)$. The intuitive meaning of this
definition is that $\epsilon(n)$ security is achieved  when the probability  that Alice and Bob do
not abort if the keys differ {\em or} that the adversary gets non-negligible information on the
final key is at most $\epsilon(n)$. In other words, except with probability $\epsilon(n)$ Alice and
Bob generate an identical key, which is unknown to the eavesdropper. It is important to note here that ITS definitions and
proofs, regard keys like $K, K^A, K^B$, as random variables, depending on an input, which is
different for different models. Keys shared finally by Alice and Bob are actually values of these
random variables. For the sake of simplicity, we ignore this difference and use capitals in what
follows.

\subsection{QKD - an ITS Cryptographic Key Agreement Primitive}
\label{sec:ITSmodels+QKD}

It is well known~\cite{Shan2,wolf:phd99} that no cryptographic method relying solely on computation
and communication over insecure communication channels can ensure ITS key establishment. In any
case additional resources given to Alice and Bob or alternatively assumptions limiting the
information available to the eavesdropper are needed to this end. ITS key agreement is possible in
a number of scenarios,  based on bounded knowledge available to the adversary, due to e.g.
intrinsic noise in the communication channel or limitations of the memory capacity of the adversary
(see ~\cite{wolf:phd99} and references therein). Alternatively ITS key agreement can also be
achieved as a consequence of the quantum nature of certain resources, e.g. a quantum communication
channel (needed for QKD), or distributed entanglement (needed for quantum
teleportation~\cite{teleport}), if such resources  are available to the legitimate parties, as these can
render unfeasible  a number of eavesdropping  activities. All methods in the discussed class
additionally assume that classical communication channels are authentic, i.e. that the  adversary
is restricted to passive eavesdropping on these channels\footnote{As pointed out below, this
additional assumption can be lifted by applying ITS message authentication schemes.}. Recently it
was found that all these methods can be formulated using a unified quantum
approach~\cite{christandl:llncs07}, based on embedding the purely classical scenarios in an
equivalent quantum framework.

Thus, from a logical point of view, QKD is just one of many methods enabling ITS key agreement.
From a more technological perspective, QKD is currently by far the least restrictive approach.
Indeed the eavesdropper is not limited by assumptions, while the additional resource required -
stable quantum communication (transmission of light quanta over optical fibers or through free
space) between Alice and Bob is already by no means a mere theoretical construction but rather an advanced
engineering practice (see e.g.~\cite{secoqc}). Simultaneously, real-time key agreement rates at
distances below 100 km reach practically usable ranges~\cite{secoqc, idsqmagiq}.

A QKD protocol generically includes two main activities: the legitimate parties communicate over a
quantum channel to get {\it correlated bit strings} and perform {\it post-processing} over the
public authentic channel to get identical secure keys or notified termination in case of technical
problems  or significant eavesdropping activity
 (see e.g.~\cite{scarani:rmp08} for details). Different methods to get correlations and different types of
post processing yield different QKD protocols. For a number of studied QKD protocols one can derive
full security proofs, which lead to explicit expressions for the information-theoretically secure key
generation rate (i.e. the length of the generated secure key per unit
time). Among the several proof techniques that have been used in the past years, the most important
ones rely on the uncertainty principle~\cite{mayers:jcm01,koashi:qp051,koashi:qp07,gottesman:qic04}, the
correspondence between entanglement distillation and classical
post-processing~\cite{lo:science283,shor:prl00}, or information-theoretic notions and in
particular smooth Renyi entropies~\cite{renner:phd05,kraus:prl05,renner:pra05}. The ultimate reason
for ITS in this case is the fact that eavesdropping attempts by the adversary on the quantum
channel, unavoidably modify quantum signals and leave signatures in form of error. The
post-processing phase allows to eradicate the knowledge acquired by moderate eavesdropping or to
recognize that information leakage is irreparable and terminate the protocol.

 Information theoretic security as introduced in Section~\ref{sec:ITS}
above, ensures in general {\it composability}~\cite{renner:llncs05}, which means that the security
of the key is guaranteed regardless of the application it is used for: if an $\varepsilon$-secure
key is used in a $\varepsilon_1$-secure task,  the composed task would be $(\varepsilon
+\varepsilon_1)$-secure. The importance of this issue for QKD was recognized only
recently~\cite{renner:llncs05}.The problem was that initial security studies adopted a security
definition which was not composable. Early security proofs defined QKD security by analogy with the
classical version of the Privacy requirement in Section~\ref{sec:ITS}: The eavesdropper, who holds
a quantum state $\rho_E$, performs the measurement $\mathcal{M}$ that maximizes her mutual
information with the key $K$. This defines the so-called accessible information
$I_{\text{acc}}(K:\rho_E) = \max_{E = \mathcal{M}(\rho_E)}I(K:E)$, and the security criterion reads
$I_{\text{acc}}(K:\rho_E) \leq \varepsilon(n)$. This was shown to be not
composable~\cite{konig:prl07}. The main problem is that this definition of security assumes that
the eavesdropper transforms her quantum state into a classical one during key agreement. In fact
she can keep her quantum state and eventually use it to break a composed task when the QKD key is
used later on. A definition that leads to composability for QKD requires a  quantum reformulation
of both ingredients (Authenticity and Privacy) of ITS. These can be embedded into a single
composable requirement~\cite{renner:llncs05} utilizing trace-norm,  $\frac{1}{2}\|\rho_{KE} -
\tau_K\otimes \rho_E\|_1 \leq \varepsilon(n)$, where $\tau_K$ is the completely mixed state on $K$.

Composability of QKD key has many implications. The most immediate one is related to relaxing the
assumption on availability of a public authentic channel. From a practical point of view this
assumption is indeed too strong. Message modification on classical channels is a simple technical
task. This would, however, allow the eavesdropper to easily mount  man-in-the middle attacks by
cutting both the classical and the quantum channels, introducing corresponding QKD quantum
technology, and carrying out two QKD protocols, one with Alice pretending that she is Bob and one
with Bob taking over the role of Alice. Fortunately, it is possible to give up the authenticity
assumption by augmenting pure QKD with a message authentication scheme, which can guarantee
integrity of classical communication with information-theoretic security. This is achieved by means
of continuous usage of secret key in classical communication. In particular, each message is sent
together with a hash value, where hashing is performed with a keyed hash function for each message
whereby the function itself is chosen from some almost universal$_2$ family of functions, which is indexed by
the secret key \cite{Weg_Car, Car_Weg}. The rate of key generation of pure QKD is higher than the
key usage for message authentication. Therefore, putting things together, QKD is an
information-theoretically secure key agreement process, which needs a fixed (small) amount of
pre-distributed initial secret key to start with. Due to composability, subsequent authentication
of communication can be performed using part of the newly generated key\footnote{It is remarkable
that the cryptographic key agreement primitive most widely used in current security practice -
namely the Diffie-Hellman key agreement protocol~\cite{diffie:ieee76}, is also prone in its pure
form to man-in-the-middle attacks and for this reason has to be augmented by additional measures.}.

\subsection{QKD Links: Performance and Application Domains}
\label{sec:QKDLinks}

Having clarified the security of QKD we turn to more practical issues like the connectivity it
allows and its typical performance.

As far as connectivity is concerned it should be noted that QKD is intrinsically a point-to-point
primitive (need for dedicated direct connection by a quantum channel, necessity of peer-to-peer key
pre-sharing), and is thus suitable for key establishment in a closed community. Further it should
be pointed out that, as a consequence of composability,  if the QKD-generated key is used for an
information-theoretically secure  communication, provided by One Time Pad (OTP) encryption {\em
together with} unconditionally secure authentication, then the composed protocol realizes an
unconditionally secure channel - a point-to-point {\it QKD link}{\footnote{A QKD link is realized
by two quantum optics and processing devices - {\em QKD devices} - usually a sender and a receiver,
deployed with Alice and Bob respectively, which generate key and optionally can perform simple key
management and ITS encryption/authentication.}, which among other tasks, can be used for key
transport as discussed in the subsequent section.

\begin{figure}
\begin{center}
\includegraphics[width = 7 cm]{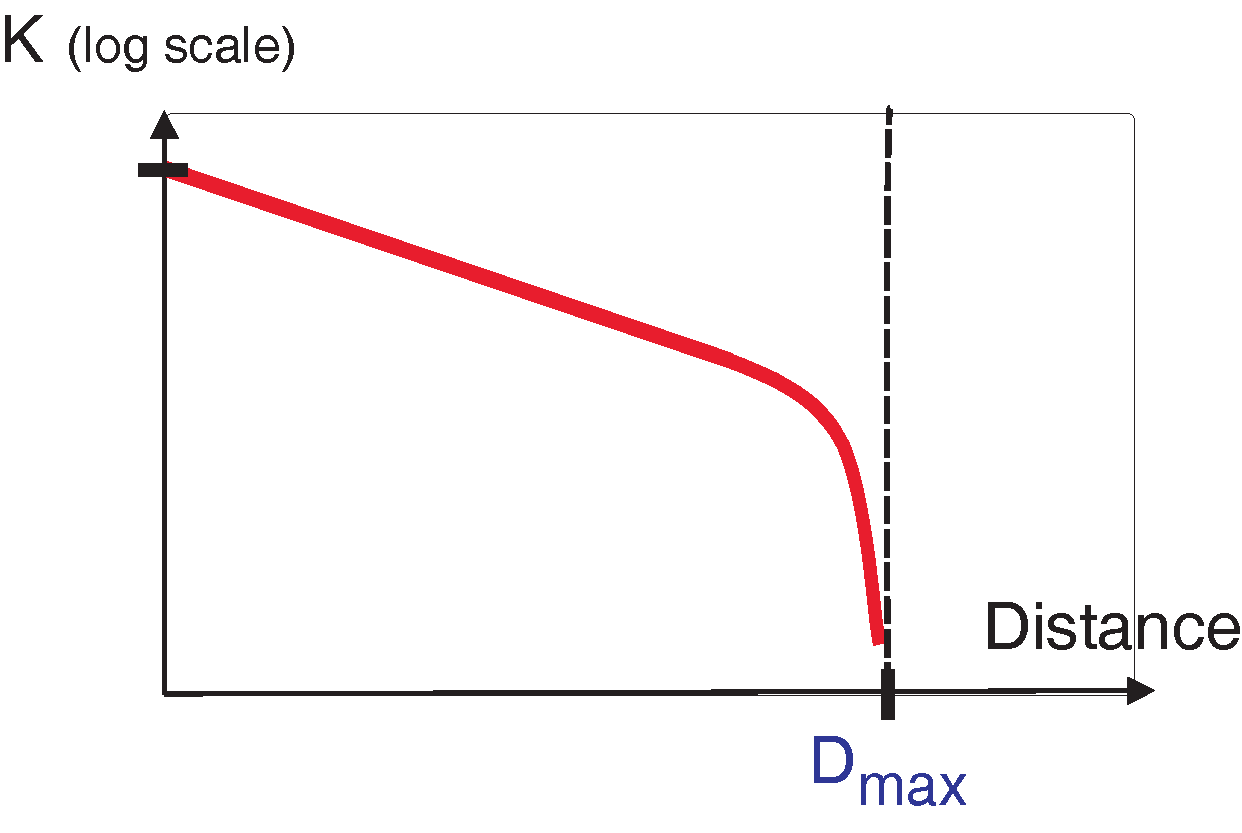}
\caption{Typical profile of the rate versus distance curve for a single QKD link.}
\label{fig:RateQKDLink}
\end{center}
\end{figure}

Performance on the other hand is given by the secret key generation rate $K(\ell)$, which is a
characteristic function of distance $\ell$ depending on the QKD protocol and the specific
implementation of a QKD link. This rate clearly varies from system to system but in general terms
it follows the curve of Fig.~\ref{fig:RateQKDLink}. As shown in this figure, the logarithm of the
rate of secret bit agreement initially falls at a given power of the channel attenuation (depending
on the implemented QKD protocol), and features an exponential drop-off at long distances. In
addition to reliability and stability, the performance of practical QKD systems is usually measured
by the maximum communication distance they can reach, $D_{\text{max}}$, and the secure key
generation rate they can achieve at a useful range. The limiting factors vary greatly for different
protocols and implementations and range from hardware-related problems such as the high dark count
rates in typical single-photon counting detectors at telecommunication wavelengths to algorithmic
issues such as the finite efficiency of error-correcting codes~\cite{scarani:rmp08}. The distance
at which direct QKD between two parties is possible is roughly limited to 100~km in optical fibers
for current systems, with a possibility of reaching up to 200~km in the next few years, while the
secret key generation rate is currently limited to a few tens or hundreds of kbit/s depending on
the distance.

It is clear from the above discussion that QKD links suffer from intrinsic limitations: they cannot
be operated over arbitrarily long distances and their use is restricted to point-to-point key
exchange/secure communication  between the two endpoints of the quantum channel. A natural question
that arises then is what could be the application field of a technology with such characteristics.
Obviously, QKD links can be directly used in an environment, in which highly secure communication
is required between two parties over a relatively short distance. If information-theoretic secure
communication is the target, it can be achieved at low rate (i.e. around 10-20~kbit/s). If broadband
secure communication is needed instead, then unconditionally secure communication is out of reach
at a reasonable cost. A highly secure point-to-point communication is still possible by combining a
pair of QKD devices with high end symmetric encryptors (typically running the AES encoding scheme).
In this case, the limit is set by the speed of encryption (around 10~Gbit/s) whereas the key is
exchanged at a rate allowed by the QKD device-pair. It should be stressed that although the overall
security offered by such {\it QKD link-encryptors} is no longer information-theoretic it greatly
exceeds the one provided by any other currently existing method. Today, several QKD-based link
encryptors are commercially available~\cite{idsqmagiq}, but their range of applications in
practical communication systems is inevitably rather limited. A better way to exploit the extremely
high security standard offered by QKD and to extend the application range  to long-distance and
multiple-user key establishment is to combine several QKD links in order to form a {\em QKD
network}. Indeed, as we will see in the next sections, a number of the aforementioned limitations
of QKD links can be overcome when it is possible to achieve QKD-based unconditionally secure key
agreement over a network~\cite{alleaume:white07,dianati:scn08}. From this perspective the
development of QKD network architectures appears  as a necessary step in order to achieve effective
integration of QKD into secure communication networks.

\section{QKD Networks}
\label{sec:QKDnetworks}

We define a {\it QKD network} as an infrastructure for ITS key establishment, which relies on
quantum resources available to the legitimate participants, while not imposing bounds on the
eavesdropping capabilities of the adversary, and allows connectivity of parties that do not share a
direct, fixed quantum channel. Optionally this infrastructure should also allow lifting the
restrictions typical for stand alone QKD links - enable ITS key establishment over long distances
(e.g. continental scale), increase and maximize the throughput capacity (the key generation rate)
and ensure robustness against denial of service attacks and technical service break-downs.

The first proof-of-principle QKD network demonstrator, the ``DARPA Quantum network'', was deployed
between Harvard University, Boston University and BBN in 2004~\cite{elliott:njp02,elliott:qp05}. A
highly integrated network demonstrator, developed within the framework of the integrated FP6
Project {\sc Secoqc}, which ensures network-wide ITS key establishment, was deployed, tested, and demonstrated
in Vienna~\cite{secoqc}.

\subsection{Types of QKD Networks}
\label{sec:QKDnettypes}

The precise notion of ITS security depends on the particular QKD network model. For this reason we
start by considering the different QKD network types. These have been known for a long time now and
have been suggested already in ~\cite{Biham}. There are two principal approaches: a) {\em quantum
channel switching paradigm} -- creating an end-to-end quantum channel (or more generally
distributing quantum resources) between Alice and Bob, or b) {\em trusted repeater paradigm} --
transport of key over many intermediate nodes, which are (at least partially) trustworthy  i.e. not
infiltrated by the eavesdropper. The two approaches are essentially different and we shall discuss
them one after the other.

\subsubsection{QKD Networks With Quantum Channel Switching}
\label{sec:QKDnetswitched}

\noindent {\bf Optically switched quantum networks:} These are networks in which some classical
optical function, like beam splitting, switching, multiplexing, demultiplexing, etc., can be
applied to the quantum signals sent over the quantum channel. The interest in such optical
networking capabilities in the context of QKD networks is that they allow going beyond the two-user
QKD. Moreover, this can be done with current technology. Active optical switching can be thus
used to allow the selective connection of any two parties with a direct quantum channel (the BBN
DARPA quantum network contained an active 2-by-2 optical switch
 that could be used to actively switch between two network topologies). Optical functions
can thus be used to realize multi-user QKD, and the intermediate sites do not need to be trusted,
since quantum signals are transmitted over a quantum channel with no interruption from one end-user
QKD device to the other one. In this sense the security analysis coincides with that for a
stand-alone QKD link. This QKD network model can however not be used to extend the distance over
which keys can be distributed. Indeed, the extra amount of optical losses introduced in the
switching devices will in reality decrease the transmission capacity of quantum channels and thus the
maximal key distribution distance. In addition, in a fully switched optical network any two parties
have to share an initial secret to be able to start the key agreement process. So, overall, this
type of networks are not scalable and thus suitable for long distance QKD.
Instead, they can be used in local or metropolitan
areas.

\noindent {\bf Quantum repeater based QKD networks:} To be able to extend the distance over which
quantum key distribution can be performed, it is necessary to fight against propagation losses that
affect the quantum signals as they travel over the quantum channel. Quantum
repeaters~\cite{briegel:prl98} can overcome the loss problem and can be used to distribute
entanglement between any two parties and therefore effectively create an  end-to-end quantum
channel across the network. A quantum repeater based network can thus be seen as a ``fully quantum''
network. As intermediate network nodes do not get any information in the process of key generation,
end-to-end unconditional security is guaranteed without the need to trust these nodes. In this
sense the security analysis also coincides with that for a stand-alone QKD link.
 Quantum repeaters however rely on elaborated quantum operations and on quantum memories
that cannot be realized with current technology. As discussed in~\cite{collins:jmo05}, quantum
nodes called quantum relays could also be used to extend the distance over which secure QKD
can be performed\footnote{Both quantum repeaters and quantum relays are devices that allow to
teleport qubits over several quantum channel segments, whereby entangled photons are distributed along the separate segments. The main difference between quantum repeaters (see~\cite{scarani:rmp08} for a simple model of a quantum repeater) and quantum relays is that while in a quantum repeater received photons are kept in quantum memories in order to bring entangled pairs from adjacent segments in correspondence, in a quantum relay one waits for the event when all photons sent along the different segments are received - i.e. none is absorbed.}. Quantum relays are
simpler to implement than quantum repeaters since they don't require quantum memories. However,
even quantum relays have not yet been technically realized. 
Moreover,  quantum relays would not allow
secure QKD over arbitrary long distances. \\

\subsubsection{Trusted Repeater QKD Networks}
\label{sec:QKDnetstrusted}

Trusted repeater QKD  networks have been discussed in various contexts since the advent of Quantum cryptography.
Below we give  a more formal definition, which in turn simplifies the subsequent security analysis of such networks.

We define a {\it QKD trusted repeater network} as an infrastructure composed of QKD links, i.e.
from a structural point of view pairs of QKD devices associated by a quantum and a classical
communication channel, each link connecting two separate locations or {\it nodes}. A QKD trusted repeater network is
then a connected graph, the vertices of which are nodes, and the edges - QKD links. 

We assume further that initial secret keys are only shared between neighboring nodes (i.e. ones
directly connected by a QKD link) and not between any arbitrary pair. This assumption ensures that
the number of initial secrets to be shared scales (for wide area networks) with the number of
network nodes and not with their square. This in turn largely simplifies the initialization of a
QKD network and the adoption of additional nodes during operation.

 QKD networks based on trusted key repeaters follow a simple principle: global key distribution is performed over
a QKD path, i.e. a one-dimensional chain of trusted repeaters connected by QKD links, establishing
a connection between two end nodes. Secret keys are forwarded, by unconditionally secure key
transport  along the QKD links of the path in a hop-by-hop fashion. (As mentioned above
unconditionally secure transport over separate QKD-links is ensured by One Time Pad encryption and
ITS authentication, both realized with a local QKD key.) End-to-end information-theoretic security
is thus obtained between the end nodes, {\em provided} that {\it all} the intermediate nodes can be
trusted, as these possess the full communicated information. The trusted nodes play thus the role of (classical) trusted repeaters.
 This architecture
 can be used to build a long-distance QKD network. The advantage of such quantum networks is that they rely
on QKD for link key establishment, which guarantees that it is impossible to compromise the network
key distribution by direct attacks on the links.

Trusted repeater QKD networks can be implemented with today's technology since the nodes are
essentially QKD devices plus classical memories and processing units placed within secure
locations. This concept had been tested in  the BBN QKD network and is also the basis of the {\sc
Secoqc} QKD network, which is exclusively based on the trusted repeater approach.

\subsection{ Security Framework and the Architecture {\sc Secoqc}}
\label{sec:SECOQC}

In the trusted repeater paradigm one can  differentiate between two basic security frameworks:

The first trust framework, already outlined above,  is highly realistic and relevant for internal networks
belonging to a spatially distributed entity such as an industrial, financial, governmental, or
military institution, the backbone of a telecommunication provider, etc. This case is the main
focus of {\sc Secoqc}. The all-nodes-trusted assumption obviously leads to a straightforward
cryptographic conclusion on the security of network connectivity. Together with the  guarantee for
an information-theoretically secure transport from node to node provided by the underlying QKD
links it ensures unconditionally secure transport between Alice and Bob. Indeed in this case the
eavesdropper is restricted to attacking the QKD links,  which at best can result in a denial of
service but not in a gain of any information on the (key) material which is securely transported.
While this argument ultimately settles the security analysis in the current model, a practical
network realization requires addressing a multitude of architectural tasks, which are of more
applied nature. These tasks include:
\begin{itemize}
\item How to design the architecture of network nodes so that they can provide a universal key
distribution mechanism, while possibly integrating heterogeneous QKD links~\cite{Poppe:ijqi08}?
(Here heterogeneity is meant in terms of the background QKD protocol and device engineering.)
 \item How to specify the peer-to-peer key transport protocols?
 \item Which particular information-theoretically secure message authentication code to select for
implementation?
\item How to design end-to-end network routing and transport
protocols, taking into account the unconditionally secure nature of the
transport~\cite{dianati:scn08}?
\item How to optimally plan the deployment of QKD networks, from a cost perspective, based on a study
of the relation of cost and topology~\cite{alleaume:qp09} ?
\end{itemize}

All of these issues have been at the core of the development work of {\sc Secoqc}. They have been
addressed by a broad interdisciplinary team, and important advances have been made in all mentioned
areas\footnote{It should be noted that currently the results are only partially publicly available,
as at present the project team continues the effort of preparing internal deliverables for final
publication. Unpublished deliverables include:
O. Maurhart, ``Q3P: A Proposal'';
M. Fitzi, ``General Authentication Framework in QKD'';
J. Bouda, et al., ``SECOQC Node Keystore Module and Crypto Engine'';
J. Bouda, et al.,``Encryption and Authentication in SECOQC''.}.
The outcome is a layered network model effectively decoupling all classical
communication as well as the network and key transport functionality from the operation of the QKD
devices. As a result, the {\sc Secoqc} network involves the ability to integrate, by using standard
interfaces, a completely heterogeneous physical layer consisting of different types of QKD devices
from multiple providers with a homogeneous network-wide end-to-end key transport layer. The project has put in operation and tested a highly integrated prototype
in the metropolitan fibre-ring of Siemens in the city of Vienna (see Fig.~\ref{fig:Secoqc_VIE} for a schematic representation]. A
public demonstration of this prototype took place October 8, 2008.

\begin{figure}
\begin{center}
\includegraphics[width = 7 cm]{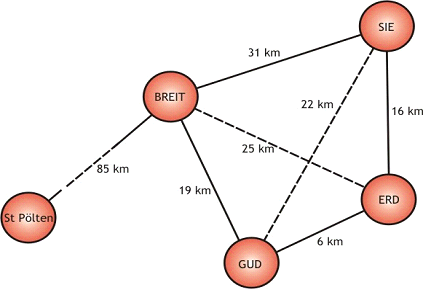}
\caption{The {\sc Secoqc} network prototype in Vienna - a sketch.} \label{fig:Secoqc_VIE}
\end{center}
\end{figure}

The second  framework type assumes that a limited number of nodes are taken over by the adversary or  {\em corrupted}.
 Obviously this framework is much more challenging from a cryptographic point of
view. It is closely related to, although distinct from, a classical problem dedicated to the study
of secure message transmission over untrusted networks~\cite{DDWY93}. In the latter model, it is assumed
that any node of the network can be taken over by the adversary but the number of corrupted nodes is upper
bounded by some threshold. Apart from the threshold, adversaries can be arbitrarily malicious or
{\em Byzantine}. Any such adversary that can take over no more than $t$ nodes is called $t$-{\em
bounded}. In Section~\ref{sec:section4}, we study the same problem for trusted repeater QKD
networks, where some nodes are corrupted and Byzantine. We discuss an essential difference
with respect to the classical case: a condition that protocols in the classical setting
should satisfy is too strong when private links between neighbouring nodes are implemented
using QKD.

It should be noted that this second framework is highly relevant for QKD networks owned by several, possibly
competing entities, and mimics realistic telecom network settings. It requires further research and
in particular addressing of all practically relevant tasks, already carried out in {\sc Secoqc} for
the case of all-node-trusted networks.

\section{Secret-Key Agreement Over a QKD Network With Corrupted Nodes}
\label{sec:section4}

In this section, we discuss privacy and authenticity of secret keys generated over a trusted
repeater QKD network with some corrupted nodes. We look at how to characterize adversaries in this
model and how to achieve security of the secret keys generated over the QKD network against these
adversaries. We compare a QKD-network approach  to the related classical problem of perfectly
secure message transmission over untrusted networks. We provide a mean by which Alice and Bob can
verify the authenticity of secret keys generated over a QKD network. This point was originally
addressed in the unpublished {\sc Secoqc} Deliverable ~\cite{DSECArch}. While the current paper has
been in preparation two preprints with similar objectives ~\cite{Sanders1, Sanders2} have been
published. The approach of the authors is similar to the one presented here, but the
techniques used to verify the authenticity of the keys are different.
The advantage of our technique
lies in its potential not only to differentiate between authentic and forged keys, but as discussed
below, to help revealing malicious parties in some scenarios.

\subsection{The Basic Setting}\label{setting}
A straightforward strategy for Alice and Bob to generate a secret key unknown to any other single
node in the network is to use two
disjoint paths.
The final key $K$ between Alice and Bob is a secret shared by these paths.

\begin{figure}[h]
 \begin{center}
  \includegraphics*[scale=0.9, width=0.9\textwidth, keepaspectratio=true]{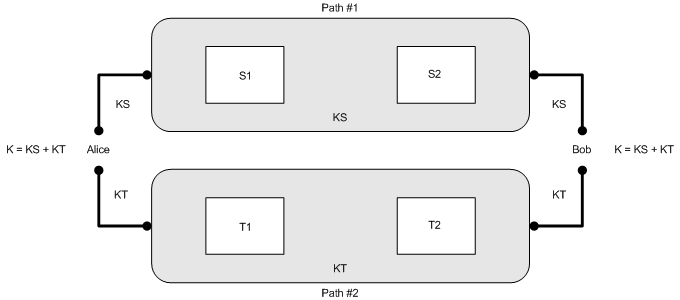}
   \caption{Example:Two paths between Alice and Bob.}\label{f2}
 \end{center}
\end{figure}

Figure~\ref{f2} shows an example where Alice and Bob will
generate a secret key $K=KS\oplus KT$ using the keys $KS$ and $KT$, which are secret-keys generated
on each path. Of course, the secret-key of each path is generated
using point-to-point QKD and the standard hop-by-hop mechanism. The secret-key
$K$ is secure and unknown to each path as long as the paths do not fully collaborate in a malicious way.
It means that $K$ is secure only if users can {\em trust} at least one path out of the two. In
general,  if Alice and Bob generate a secret-key $K$ from $t$ paths then $K$ will be secure unless
all $t$ paths are dishonest and collaborate. We denote by $\{K_i\}_{i=1}^t$ the set of all $t$
intermediary keys of length $n$ and we let $K := \bigoplus_{i=1}^t K_i$ where '$\bigoplus $' denotes the bitwise exclusive-or.

Notice that the point-of-view described above is relevant in practice when each path is owned by a
single entity. In this case, nodes along a path do not have a life on their own but are rather
representative of a single authority. When only one node misbehaves along a path, the entire path
becomes dishonest. In this setting, paths are rather static since they correspond to physical
authorities\footnote{This basic model was introduced in one of the first cryptography deliverables
of {\sc Secoqc}~\cite{DSEC}.}.

\subsection{Private Transmission Over Classical Untrusted Networks}
\label{classical_transmission}

We have informally discussed classical secure message transmission protocols in
Section~\ref{sec:KEP_QKD}. A little bit more formally, perfectly secure message transmission
protocols against $t$-bounded adversaries, i.e. adversaries controlling no-more than $t$ nodes,
should satisfy the two following properties:
\begin{description}
\item[Guaranteed Delivery:] No $t$-bounded adversary can prevent Alice's message
      to reach Bob, and
\item[Privacy:] No $t$-bounded adversary has access to more than a negligible amount
      of information about the message sent by Alice.
\end{description}

In this model, Dolev, Dwork, Waarts, and Yung\cite{DDWY93} have shown the following with respect to
one-way communication links. Links are said to be one-way if the connectivity graph of the network
is a directed graph.
\begin{enumerate}
\item When all communication links are one-way without feedback, they
       show that it is necessary and sufficient to have $3t+1$ vertex
       disjoint directed paths from Alice to Bob. For any two nodes
       to be able to communicate privately, the network graph must be
       $3t+1$ connected (sufficient and necessary condition).
\item When all communication links (edges in the graph) are two-way, they
       show that $2t+1$ vertex disjoint paths are necessary and sufficient
       for Alice and Bob. For any two nodes to be able to communicate privately,
       the network graph must be  $2t+1$ connected (sufficient and necessary).
\end{enumerate}
Notice that privacy is more demanding than reliability since in order to have a private
communication it is necessary to have a reliable one! More precisely, if in a point-to-point
network an adversary can hack up to $t$ nodes then a $t+1$ vertex disjoint directed graph is
sufficient for reliable communication alone.


This model has been generalized by Desmedt and Wang\cite{DW02} where they consider the possibility
of using some feedback channels. Feedback channels become possible when the connectivity graph of
the network is not one-way directed outside all nodes.  When $u$ feedback channels are vertex
disjoint from the forward channels they show that:
\begin{enumerate}
\item When there are $2(t-u)+1\geq t+1$  directed disjoint paths from Alice to
       Bob, private message transmission is possible against $t$-bounded
       adversaries where there are $u$ directed node disjoint paths from
       Bob to Alice. As mentioned above,
       these $u$ paths must also be node disjoint from
       the $2(t-u)+1$ paths from Alice to Bob.
\item When there are $3t+1-u\geq 2t+1$  directed disjoint paths from Alice to
       Bob and $u$ directed paths from Bob to Alice
       (where as before paths from Alice to Bob and paths from Bob to Alice
        are node disjoint) it is possible to have
       private message transmission against $t$-bounded
       adversaries.
\end{enumerate}
These results were improved in \cite{DW06} by giving necessary and sufficient conditions for
private message transmission with feedback. Again for the case where the feedback channels are vertex
disjoint from the forward channels, we have:

\noindent{\bf Theorem 1.}[\cite{DW06}]\label{thdw06} {\em Assume there are $u$ directed node
disjoint paths from Bob to Alice, vertex disjoint from the forward channels. Then a necessary and
sufficient condition for private message transmission from Alice to Bob against any $t$-bounded
adversaries is that there are $\max{\{3t+1-2u,2t+1\}}$ directed node disjoint paths from Alice and
Bob.}

Notice that all these results put serious restrictions on the number of available disjoint paths
between the two parties who want to communicate privately. Without feedback, in order to protect
against a mere 3 corrupted nodes, Alice and Bob must be able to communicate through 10 disjoint
paths while if all connections are two-ways then 7 paths are required.

\subsection{Differences with the QKD Setting}\label{diff}

In this section we quickly and roughly discuss the main differences between the classical and the
QKD (trusted repeater) setting for private communication on untrusted networks.

The most obvious difference between the two settings is that while in the classical case messages
are transmitted, a QKD network  is mainly concerned with key distribution. This difference is only
cosmetic. It is easy to see that private message transmission implies the ability to distribute
secret-keys and that the ability to distribute key implies the ability to send private messages. In
other words, the functionalities achieved in both settings are equivalent.

Like for private classical message transmission, privacy of secret-keys generated through a QKD
network  can only be guaranteed if different paths do not overlap. If a corrupted node $N^*$ is
part of all quantum paths between Alice and Bob then no private communication (or key) can possibly
be established. Therefore and unless nodes taking place in more than one paths are incorruptible,
we can focus on network architectures with non-overlapping
paths. 

While for classical private message transmissions point-to-point private communication is assumed
between any neighboring nodes, in a QKD network no such assumption is required since private
point-to-point communication is provided by QKD. It follows that all private message transmissions
protocols and in particular the ones of \cite{DDWY93,DW02,DW06} can be implemented using QKD to
provide private point-to-point communications between neighboring nodes. Using these classical
constructions would allow for key distribution and private communication against more general
network architectures than the one depicted in Fig.~\ref{f2}. Moreover, when QKD is used to
implement private point-to-point communication in the constructions of \cite{DDWY93,DW02} for
instance, $t$-bounded adversaries can in addition to controling any $t$ nodes, eavesdrop the classical
communication between any other pair of nodes. If in addition the adversary eavesdrops the quantum
channel then although it becomes possible to implement a denial of service attack\footnote{Too much eavesdropping on the quantum channel will cause two neighbouring nodes to abort
the key generation.}
 no information
on a secret key successfully generated can be obtained.

Then, how come that the situation depicted in Fig.~\ref{f2} allows for Alice and Bob to agree upon a
secret against any $1$-bounded adversaries while there are only $2$ disjoint paths in the network?
This seems to do better than the necessary $2t+1$ paths of \cite{DDWY93}. The answer is that
in the situation depicted, Alice's and Bob's keys were not required to be identical
but only to be both unknown to the
adversary. It is straightforward for one corrupted node to prevent Alice and Bob from agreeing on
an identical key. Moreover, Alice and Bob will not be able to detect that they do not share a
private key unless they already share an authentication key used to establish the correctness of a
newly  generated secret key. Unlike for the classical case described in
Sect.~\ref{classical_transmission}, the rough setting described above 
does not
address the problem of {\em guaranteed delivery}. This may have important consequences for the
security of the architecture. Such weakness is not a desirable property for any network
architecture providing privacy. However, {\em guaranteed delivery} seems to be asking for too much
since QKD never guarantees successful key generation; a denial of service
attack is always possible in principle.

This circumstance calls for a slightly weaker delivery condition in the QKD-network case in
comparison to the fully classical setting.  Instead of {\em guaranteed delivery}, it is more
appropriate to require either an authentic delivery to both parties (the keys of
Alice and Bob coincide and they know it) or a notification of network failure. More formally we require a
delivery condition which is analogous to an ITS end-to-end key establishment between two arbitrary
nodes (Alice and Bob) over the network.
\begin{description}
\item[Authenticity:] Any two parties Alice and Bob can send {\em
       classical} messages
       between them in a way that will either guarantee delivery
       and therefore $K^A=K^B$ or lead to a notification of a network failure.  This is
       weaker than the {\em guaranteed delivery criterion} discussed in
       Section~\ref{classical_transmission}.
\item[Privacy:] No 
adversary has information about neither $K^A$ nor $K^B$ generated
  by Alice and Bob during key generation. In particular, when
  $K^A=K^B$ the adversary has no information about the secret key.
\end{description}

Notice that for the sake of clarity we  have deliberately simplified the definition by omitting the
$\epsilon(n)$ notation although we keep it in mind.

\subsection{Achieving Privacy and Authenticity in QKD-Networks}\label{sec}

In order to achieve both {\em authenticity} and {\em privacy} in a QKD-network, it must satisfy
conditions similar to the ones we have seen in Theorem 1. In particular, two parties who want to
exchange a secret-key do not in general share an authentication key. It follows that testing the
authenticity of a newly generated secret-key must be performed by transferring an authentication
tag through a network where some nodes are corrupt. We shall see in the following that
authenticity is guaranteed against any ($\ell-1$)--bounded adversary if there are $\ell$  disjoint
paths. Security of the resulting secret-key is also guaranteed against ($\ell-1$)--bounded
adversaries according the security criterion of Sect.~\ref{diff} while it is guaranteed against any
($\ell-2$)--bounded adversaries according a more stringent privacy criterion that we introduce in
Sect.~\ref{priv}.
This is in any case better than  
the constructions discussed in Sect.~\ref{classical_transmission} that, while satisfying the
stronger security notion of {\em guaranteed delivery}, are secure against $t$--bounded adversaries
only if $2t+1$ disjoint channels are available.

Let us get back to authenticity and privacy of the secret-keys generated in a  QKD-network.

\subsubsection{Privacy}\label{priv} What do we mean when we say that a key obtained by Alice and Bob
is private? It is certainly not completely private since keys are also known to an adversary
controlling all paths. Even if one path is not under the control of the adversary, Alice and Bob do
not want their keys to be known by any node along a honest path. In other words, trusted nodes
should never get any information about secret keys generated through them\footnote{Consider an honest path between Alice and Bob  belonging e.g. to an organization related to them. It could happen that Alice and Bob want to share sensitive information about the organization of this very path. Even if by definition the path is honest and always properly executes the communication protocol, it could still be curious. Obviously in many cases, as the one just outlined, Alice and Bob would prefer that their communication remains private, i.e. unknown to the path.}.

Remember how secret keys are generated  when Alice and Bob are connected through $\ell$ disjoint
paths $P_1,P_2,\ldots,P_\ell$. Let $K^A_i$ and $K^B_i$ be Alice's and Bob's secret key respectively
obtained from path $P_i$, $1\leq i \leq \ell$ using QKD between neighbors. Alice and Bob then set
their secret key as:
\[ K^A := \bigoplus_{i=1}^\ell K^A_i \mbox{ and } K^B :=\bigoplus_{i=1}^{\ell} K^B_i.\]
When no adversary acts actively, the key generation is such that $K^A_i=K^B_i$ for all $1\leq i\leq
\ell$ and therefore $K^A=K^B$.

    Notice that any $t$--bounded adversary ${\cal A}$ can only learn keys $K^A_i$ and/or $K^B_i$ if
$P_i$ is under the control of ${\cal A}$. This is guaranteed by the privacy of QKD between
neighboring nodes. Let $S_{\cal A}\subseteq \{P_1,\ldots,P_{\ell}\}$ be the set of paths under the
control of ${\cal A}$. Since ${\cal A}$ is $t$--bounded we have that $|S_{\cal A}|\leq t$. By
construction, any $K^A_i$ and $K^{B}_i$ with $P_i\notin S_{\cal A}$ is completely unknown to ${\cal
A}$. It follows that both final keys $K^A$ and $K^B$ are unknown to ${\cal A}$  as soon as
$\ell>t$. Let us be more precise. Keys of length $n$ generated by QKD between honest neighbors are
guaranteed to be $\epsilon(n)$--private against any third party. (In Section~\ref{sec:KEP_QKD} we
have already pointed out that a key $K$ is $\epsilon(n)$--private if given the state of the
adversary, $K$ is $\epsilon(n)$--indistinguishable from a random $n$--bit string.)
Keys $K^A$ and $K^B$ must therefore be $\epsilon(n)$--private against any ($\ell-1$)--bounded adversary ${\cal A}$. In other words,\\

\noindent
{\bf Lemma 1.} {\em Let $K^A=\oplus_{i=1}^{\ell} K^A_i$ and $K^B=\oplus_{i=1}^{\ell}K^B_i$ be such that  $\{K^A_i\}_{i=1}^\ell$ and $\{K^B_i\}_{i=1}^{\ell}$
have been generated through disjoint paths $P_1,\ldots,P_\ell$ where $(K^A_i,K^B_i)$ is $\epsilon(n)$--private and satisfies $K^A_i=K^B_i$ when $P_i$ is a honest path.
Then, $(K^A,K^B)$ is $\epsilon(n)$--private against any ($\ell-1$)--bounded adversary but not necessarily such that $K^A=K^B$.}\\

As stated in the above lemma, ${\cal A}$ can certainly prevent Alice and Bob from generating
$K^A=K^B$. It suffices for one adversarial node to make its neighboring node to believe they share
a key while in fact they don't. It is sufficient for ${\cal A}$ to send classical messages
different from what is expected in order for $K^A\neq K^B$. Although such attack will not allow
${\cal A}$ to learn anything about $K^A$ and $K^B$, it ensures that no secure transmission can take
place between Alice and Bob even though they are not aware of this fact.

The authenticity of $K^A$ and $K^B$ should therefore be checked upon all new key generations.

Another important point regarding privacy is the following. Suppose an adversary controls $\ell-1$
paths $P_1,\ldots,P_{\ell-1}$. The honest path $P_\ell$ without behaving dishonestly could be able to
determine Alice and Bob's secret key if the adversary decided the {\em broadcast}
$\{K^A_i\}_{i=1}^{\ell-1}$. Moreover, a dishonest
path could be tempted to publish all information they gather in order to implement
a denial of service attack. Publishing this information means that honest-but-curious paths
would be able to decipher any communication between the end users. This could deter users to
use their keys. It would therefore be desirable to enhance privacy against honest-but-curious paths this way.
\begin{description}
\item[Privacy Against Honest-but-Curious Paths:] Privacy is guaranteed against honestly behaving
paths that happen to learn information from adversarial paths posting their secret information.
Privacy in this case can be enforced simply by having at least $2$ honest (but maybe curious)
disjoint paths.
\end{description}

Clearly, if two paths are honest (but curious) and even in the case when the adversary publishes
everything she knows, none of the curious but otherwise honest path learns anything about the
secret key. This follows since the secret key is shared among two honest parties who therefore
never publish any of their private information.

\subsubsection{Authenticity}\label{auth}

As mentioned above, in  a QKD-network it is desirable to pre-distribute authentication keys only for point-to-point
connections. This choice limits drastically the complexity of initial key distribution phase
required before key material can start being generated. It follows that in this model, Alice and
Bob do not necessarily have an authentic channel they could use for testing the authenticity of a
newly generated key. As discussed in the previous subsection, it is important for any pair of users
to be able to guarantee the authenticity of a newly generated secret key even though they don't
have access to an authentic channel between them.

It follows that authentication tags must be sent through channels that may be under the control of
the adversary. The key authentication process must guarantee that Alice concludes that $K^A=K^B$ if
and only if Bob concludes that $K^B=K^A$. Clearly, we also want that when $K^A=K^B$ Alice and Bob
identify this case with success.

There are different methods to get this working. Suppose Alice and Bob have generated keys $K^A$
and $K^B$ respectively where both are $n$--bit strings. They now want to establish the authenticity
of their respective key. This process should work when any $t$ paths out of $\ell$ disjoint paths
are under the control of the adversary ${\cal A}$. That is, the authenticity or non-authenticity of
a secret key should be guaranteed against $t$-bounded adversaries.

Remember from Section~\ref{priv} that over $\ell$ disjoint paths, no $t$-bounded adversary for
$t<\ell$ gets to know anything about both $K^A$ and $K^B$. It suggests to use part of $K^A$ and
$K^B$ to authenticate $K^A$ and $K^B$ through the $\ell$ disjoint paths from which each partial
keys $\{K^A_i\}_{i=1}^\ell$ and $\{K^B_i\}_{i=1}^\ell$ has been generated.

This can be done as shown in the following example.

\subsection{Example of a Simple QKD-Network}\label{ex}

For simplicity, let us get back to the example of Figure~\ref{f2} where Alice and Bob use  two
non-overlapping paths $P_1$ and $P_2$ to perform a key exchange.  In this case, the secret key
$K^A$ and $K^B$ must be authenticated and acknowledged even when $P_1$ or $P_2$ is under the
control of the adversary. From privacy however, when Alice and Bob happen to have $K^A=K^B$
they in fact have an authenticated channel between them.
Assume that $MAC_{\kappa}(M)$ is the tag of a message authentication code for message $M$ under
secret key $\kappa$. Suppose also that $MAC_{\kappa}$ can be used to authenticate two messages
securely against impersonation even if both tags have been computed with the same key $\kappa$.

One simple way to proceed in order to verify that $K^A=K^B$ in this scenario is as follows.

\begin{itemize}
\item Alice and
Bob  pick the $s$ first bits
of their respective keys denoted
by $\kappa_A = (K^A)_{1\ldots s}$ and  $\kappa_B = (K^B)_{1\ldots s}$.
Alice and Bob set $K_A = (K^A)_{s+1\ldots n}$ and $K_B = (K^B)_{s+1\ldots n}$
respectively.
\item Alice picks random $\lambda_v \in_R
\{0,1\}^{n-s}$, for $1\leq v\leq m$ where $m$ and $s$ are security parameters. Alice then sets
$M_A := (\lambda_v, \lambda_v \odot K_A)_{1\leq v\leq m}\in\{0,1\}^{m(n-s+1)}$
 where `$\odot$` denotes the inner product mod $2$.
\item Alice sends $M_A$ to Bob
together an authentication  tag:
\[ T := MAC_{\kappa_A}(M_A) := MAC_{\kappa_A}((\lambda_v, \lambda_v \odot K_A)_{1\leq v\leq m}).\]

The transmission of $(M_A,T)$ is made through paths $P_1$ and $P_2$ (that is through all paths).

\item
Let $M_1$ and $M_2$ be the message received from path $P_1$ and $P_2$ respectively. Bob, upon
reception of $M_1=M_2=((\lambda'_v, r_v)_{1\leq v\leq  m},T)$, verifies  that
\begin{equation}\label{mac}
 T \stackrel{?}{:=} MAC_{\kappa_B}((\lambda'_v, \lambda'_v \odot K_B)_{1\leq v\leq m}).
\end{equation}
\end{itemize}

Since for $\lambda_v$ chosen at random in $\{0,1\}^n$, when $K_A\neq K_B$ \[ \Pr{\left( \lambda_v
\odot K_A \neq \lambda_v\odot K_B\right)} = \frac{1}{2},\] it follows that if $K_A\neq K_B$ then
Bob will observe at least one $1\leq v^*\leq m$ such that $\lambda'_{v^*} \odot K_B \neq r_{v^*}$
except with probability $2^{-m}$. When
Bob verifies that $T$ is well formed and that for each $1\leq v\leq m$, $\lambda'_{v} \odot K^B =
r_{v}$ then he outputs $res := {\sc ok}$. Notice that when $M_1\neq M_2$ and one $M_b, b\in\{1,2\}$
is a properly authenticated transmission of $M_A$ then  Bob can still set $res := {\sc ok}$ in
addition to identify that path $P_{3-b}$ is dishonest. Otherwise, when $M_1$ and $M_2$ are not
properly authenticated with key $\kappa_B$, Bob outputs $res := {\sc fail}$. Bob also outputs
$res := {\sc fail}$ if he finds at least one $v^*$, $1\leq v^*\leq m$ such that $\lambda'_{v^*} \odot K_B \neq r_{v^*}$.
Bob then authenticates
the output $res$ by computing the tag \[ T' := MAC_{\kappa_B}(res).\] As for Alice's transmission,
Bob sends $M_B:=(res,T')$ through each path $P_1$ and $P_2$. Alice receives $M'_1$ and $M'_2$ from
$P_1$ and $P_2$ respectively. If neither $M'_1$ nor $M'_2$ is properly authenticated with session
key $\kappa_A$ then Alice concludes that $K_A\neq K_B$. If Bob has determined that $K_A=K_B$ then
$(res,T')$ is a properly authenticated message with key $\kappa_A$ and can therefore be checked by
Alice. Since at least one of $P_1$ or $P_2$ is honest, Alice will get Bob's message $M_B$ in $M'_1$
or $M'_2$ (or both!) and this can be checked since messages are authenticated. This means that if
either $P_1$ or $P_2$ misbehaves during the transmission of $M_B$ then Alice will be able to
identify the dishonest path.  It follows that when  Bob concludes $K_A=K_B$  then Alice reaches the
same conclusion. Moreover, when $K_A\neq K_B$ Alice also determines it since no message among
$M'_1$ and $M'_2$ is properly authenticated since $\kappa_A \neq \kappa_B$ and since $MAC$ is a
secure authentication scheme. Notice that no adversary (controlling one path in this case and
$\ell-1$ paths when there are $\ell$ disjoint paths) can forge an authenticated message since from
the discussion of Section~\ref{priv}, the adversary has {\em no information} about neither $K^A$
nor $K^B$ and therefore neither $\kappa_A$ nor $\kappa_B$.

\subsection{Providing Secret Key Authenticity}\label{method}

In this section, we describe how Alice and Bob can determine the authenticity of a
newly generated secret key given that they use a secret-key generation
over $\ell>t$ disjoint paths $P_1,\ldots,P_\ell$. We assume that $MAC_{\kappa}(M)$ denotes the
authentication tag of message $M$ using key $\kappa\in\{0,1\}^s$. For simplicity, we also assume
that $MAC$ is secure against impersonation even given two messages-tags pairs authenticated with
the same key. These schemes are easy to construct and we will discuss this point in
Section~\ref{mac}. In the following, we denote by $p_{\tt im}$ the probability of successful
impersonation even after having seen two pairs message-tag.

Now, we have to make an assumption about the behavior of honest paths. When Alice sends a message
$M$ to Bob through honest path $P$, $M$ is sent from node-to-node until it reaches Bob. Each
transmission between neighboring nodes $N_i$ and $N_{i+1}$ is authenticated. An adversary however
could, in theory, prevent $M$ from reaching $N_{i+1}$. If this is the case, Alice could be unaware
of Bob's status since she never received his last message. This suggests to consider
quantum networks where
\begin{center}
\begin{equation}\label{assumption}
\begin{minipage}[c]{0.80\linewidth}
{\em Any classical message $M$ from neighboring nodes $N_{i}$ to $N_{i+1}$ along a honest path will eventually reach $N_{i+1}$. }\\
\end{minipage}
\end{equation}
\end{center}
The reason why this assumption does not seem to be too strong is the following. Any neighboring
nodes $N_i$ and $N_{i+1}$ share an authentication key. They can therefore use any network
connecting them in order to transmit authenticated information. Although possible, it is unlikely
that an adversary can succeed in preventing $N_i$ and $N_{i+1}$ from communicating forever. In
practice, the internet can almost be considered as a network where information between parties is
always delivered. Notice also that if messages between neighboring nodes cannot be delivered then
the privacy of keys will never be compromised but only the agreement between the end users upon
whether their respective keys are identical is.

The following procedure generalizes the approach described in Section~\ref{ex} to the case where
the number of channels is arbitrary. We shall prove in the following that this scheme provides a
secure way of verifying the authenticity of the secret keys under assumption (\ref{assumption}).

\begin{enumerate}
\item {\em Public information:} $|K^A|=|K^B|=n$, $m<n-s$ (security parameter for the probabilistic
test of $K^A=K^B$), and $s<n$ (the key size for a public MAC), and $\ell\geq 2$ (number of disjoint paths).
\item Alice sets $\kappa_A := (K^A)_{1\ldots s}$, and $K_A := (K^A)_{s+1\ldots n}$ and similarly
Bob sets $\kappa_B := (K^B)_{1\ldots s}$,  and $K_B := (K^B)_{s+1\ldots n}$. \item Alice picks
random $n-s$-bit strings $\lambda_v\in_R \{0,1\}^{n-s}$ for $v=1\ldots m$. She forms the
$m(n-s+1)$-bit string $M_A := (\lambda_1||\lambda_1 \odot K_A,\ldots,\lambda_m||\lambda_m\odot
K_A)$ where `$||$` denotes string concatenation. She computes the tag $T$ associated to $M_A$: \[ T
:= MAC_{\kappa_A}(M_A). \] \item Alice sends copies of $(M_A,T)$ to Bob through each path
$P_1,\ldots, P_\ell$. Along each path $P_i$,
      $(M_A,T)$ is transmitted from point to point in an authentic way using the authentication key shared between neighbors.
\item Bob collects all messages $(M_1,\ldots,M_{\ell})$ received from paths $P_1,\ldots,P_\ell$.
Bob locates one $1\leq h \leq \ell$ such that $M_{h}=(M^h_A,T^h)$ and $T^h =
MAC_{\kappa_B}(M^h_A)$. If such $h$ cannot be found then Bob sets ${\tt result}=0$. Otherwise, Bob
verifies that for $M^h_A = (\lambda'_1||\lambda'_1\odot K_B,\ldots, \lambda'_m ||\lambda'_m\odot
K_B)$. If this is not the case then ${\tt result}=0$ otherwise Bob sets ${\tt result}=1$. \item Bob
sends $({\tt result},T')$ where $T' = MAC_{\kappa_B}({\tt result})$ through each path $P_i$ the
same way as Alice did it for $(M_A,T)$. Let $M'_1,\ldots,M'_\ell$ be all messages received through
each path $P_1,\ldots,P_\ell$. \item Alice verifies  that for some $1\leq h'\leq \ell$, $M_{h'} =
(r'', T'')$ where $T''=MAC_{\kappa_A}(r'')$ and $r'' \in \{0,1\}$. If it is not the case then she
sets ${\tt result'}=0$, otherwise she sets ${\tt result'}=r''$. \item {\em final step:}
    \begin{itemize}
     \item If ${\tt result}=1$ Bob accepts key $K_B$ as a newly authenticated secret key with Alice. Otherwise, $K_B$ is discarded.
     \item If ${\tt result'}=1$ then Alice accepts key $K_A$ as a newly authenticated secret key with Bob.  Otherwise, $K_A$ is discarded.
    \end{itemize}
\end{enumerate}

Notice that it is important that at least one copy of both $M_A$ and ${\tt result}$ eventually
reaches its intended receiver. Otherwise, Bob after detecting $K_A=K_B$ could leave Alice unaware
of this fact if the adversary prevent message $M_B$  from ever reaching Alice untampered with. In
this case Alice would conclude that Bob observed $K_A\neq K_B$. Under assumption \ref{assumption}
however, it is guaranteed that Alice and Bob agree on the output of the authentication process.
Moreover, when $K_A=K_B$ is agreed upon by Alice and Bob then
$K_A=K_B$ except with vanishingly small probability. Before proving this, let's denote by $\delta_{K,K'}$ the function that returns 1 if $K=K'$ and $0$ otherwise where $K$ and $K'$ are bit strings. We're now ready to prove the correctness of the key authentication process.\\

\noindent{\bf Lemma 2.} {\em Assume Alice and Bob have generated $\epsilon(n)$--private secret keys
$K^A\in\{0,1\}^{n}$ and $K^B\in\{0,1\}^n$ through  disjoint paths $P_1,\ldots,P_{\ell-1}$, and
$P_{\ell}$ under assumption (\ref{assumption}). The secret key authentication process results in}
\begin{equation}\label{the}
 \Pr{({\tt result}={\tt result'}=\delta_{K_A,K_B})} \geq (1-\epsilon(n))(1-2^{-m})(1-p_{\tt im})^{2\ell-2}.
\end{equation}

\noindent{\em Proof.} Suppose first that $K^A$ and $K^B$ are uniform and random from any
$t$--bounded adversary. This happens with
probability at least $1-\epsilon(n)$ by definition of $\epsilon(n)$--privacy.

Second, suppose that $K^A=K^B$. By assumption (\ref{assumption}), there exists at least one $h$
such that $M_h=(M_A^h,T^h)$ and $T^h=MAC_{\kappa_B}(M^h_A)$. The probability $p_{{\tt no-del}}$
that $M_h=(M_A^h,T^h)\neq (M_A, T)$ is no more than the probability that one impersonation of
adversary ${\cal A}$ succeeds. By definition of the impersonation probability $p_{{\tt im}}$ for
the $MAC$ scheme, we have
\begin{equation}\label{nodel}
 p_{{\tt no-del}} \leq 1-(1-p_{{\tt im}})^{\ell-1}
\end{equation}
since the adversary is ($\ell-1$)--bounded. Upon successful delivery of $(M_A,T)$, Bob always sets
${\tt result}=1$ since the equality test never gets it wrong when $K^A=K^B$. Bob's message $({\tt
result},T')$ to Alice will also be received as such by the same probability $p_{{\tt no-del}}$ as
defined in (\ref{nodel}). It follows that,
\begin{equation}\label{equal}
 \Pr{({\tt result}={\tt result'}=1|K_A=K_B)} \geq (1-p_{{\tt im}})^{2\ell-2}.
\end{equation}

Third, assume that $K^A\neq K^B$. As when $K^A=K^B$, Bob will successfully receive $(M_A, T)$
except with probability $p_{{\tt no-del}}$. Either Bob manages to find $h$ such that $M_h=(M_A^h,
T^h)$ and $T^h=MAC_{\kappa_B}(M^h_A)$ or not. If not then by assumption~\ref{assumption} it follows
that $K^A\neq K^B$ and Bob will set ${\tt result}=0$. By the argument that lead to (\ref{equal}) we
have,
\begin{equation}\label{sub}
\begin{split}
 \Pr{({\tt result}={\tt result'}=0|K_A\neq K_B,(\forall h)[T^h\neq MAC_{\kappa_B}(M^h_A)])}\\\geq (1-p_{\tt im})^{2\ell-2}.
\end{split}
\end{equation}
Finally, suppose that there exists $h$ such that $M_h=(M_A^h, T^h)$ and
$T^h=MAC_{\kappa_B}(M^h_A)$. Except with probability at most $p_{{\tt no-del}}$ it is the case that
$(M_A^h,T^h)=(M_A,T)$. In particular, it means that Bob knows $\lambda_v \odot K_A$ and $\lambda_v$
for all $1\leq v\leq m$. Provided $K_A\neq K_B$, Bob will determine this fact except with
probability $p_{\text{error}}\leq 2^{-m}$. Using the same argument as the one that lead to
(\ref{equal}), we get
\begin{equation}\label{nonequal}
 \begin{split}
 \Pr{({\tt result}={\tt result'}=0|K_A\neq K_B,
  T^h=MAC_{\kappa_B}(M^h_A))}\\ \geq (1-2^{-m})(1-p_{\tt im})^{2\ell-2}.
 \end{split}
\end{equation}

Putting (\ref{equal}), (\ref{sub}), and (\ref{nonequal}) together leads to
(\ref{the}) after an extra multiplicative factor of $(1-\epsilon(n))$ is added since
the analysis above holds when $K^A$ and $K^B$ are uniform and random from the adversary's point of
view which happens with probability $1-\epsilon(n)$. $\Box$

\subsection{Recovery from Privacy Losses}\label{lost}

Lemma 2 tells us that results of both parties coincide and represent the answer to the
question $K_A\stackrel{?}{=}K_B$ except with negligible probability. What the theorem does not tell
us is how much privacy is preserved by the authentication process. In particular, all parity checks
$\{\lambda_v\odot K_A\}_{v=1}^{m}$ leaks $m$ bit of information about the key to the adversary. How
do Alice and Bob get rid of this extra leakage? One way to do it would be to use privacy
amplification but this seems an overkill. Using the interpretation of $\epsilon(n)$--privacy, Alice
and Bob can do better without the need to agree upon a random hashing function or to communicate.
That is, privacy amplification can be performed by a deterministic process.

Let us describe what Alice would do to remove the information on $K_A\in\{0,1\}^{n-s}$ leaked by
the parity check sent to Bob during the authentication process. Suppose furthermore that the
original $K^A$ was $\epsilon(n)$--private toward any ($\ell-1$)--bounded adversary as guaranteed by
Lemma 1. The following procedure gets rid of all extra information leaked during the
key-authentication process provided it was successful.

\begin{enumerate}
\item Let $\{\lambda_v\}_{v=1}^m$ be the set of parity checks sent by Alice to Bob during the  key
authentication process. Suppose the process was successful (i.e. ${\tt result'}=1$) was initially
run upon  an $\epsilon(n)$--private key $K^A$. The following produces a final
$\epsilon(n)$--private secret key $K^*_A$. \item Set the set of trashed bits to be initially empty
$\mho := \emptyset$. \item For each $1\leq v \leq m$ do:
  \begin{enumerate}
    \item Find the smallest $1\leq i \leq n-s$ such that $\lambda_{v,i}=1$ such that $i\notin \mho$.
    \item If such $i$ exists then $\mho := \mho \cup \{i\}$ otherwise do nothing.\label{rem}
  \end{enumerate}
\item Set $K^*_A := K_A- \mho$ (i.e. in other words, we remove from $K_A$ all positions
$i\in\mho$).
\end{enumerate}
Bob can certainly perform the exact same procedure on his side since he knows $\{\lambda_v\}_{v=1}^m$ upon ${\tt result}=1$. Clearly, if $K_A=K_B$ then $K^*=K^*_A=K^*_B$ and $K^*$ is shorter than $K_A$ and $K_B$ by at most $m$-bits. This is optimal since $m$ bits of information about $K_A$ (and $K_B$!) are disclosed by the key authentication process.\\

\noindent {\bf Lemma 3.} {\em The deterministic privacy amplification procedure described above when run upon key $K=K_A=K_B\in\{0,1\}^{n-s}$ that were initially (before the parity checks were revealed) $\epsilon(n)$--private produces an $\epsilon(n)$--private final secret-key $K^*\in\{0,1\}^{n-s-m}$.}\\

\noindent{\em Proof.} Let $K=K_A=K_B$ be the keys agreed upon after the key authentication process
was successful. Suppose that $K$ is really uniform and random from the adversary's point of view.
Then, each time a new bit at position $i$ is removed at Step~\ref{rem} when inspecting $\lambda_v$
all bits in the remaining positions remain uniformly distributed given $\lambda_1\odot K,\ldots,
\lambda_v\odot K$. If such a position $i$ cannot be found then  obviously $\lambda_v\odot K$ does
not leak any extra information about $K^*$ since all bits (which are uniform and random) involved
in the new parity check have already been removed from $K_A$.

In fact $K$ is not uniform and random from the adversary's point of view but rather
$\epsilon(n)$--private. However, except with probability $\epsilon(n)$, $K$ really behaves like a
uniform and random key from the adversary's perspective. It follows that except with probability
$\epsilon(n)$, the deterministic privacy amplification process produces a uniform and random key
$K^*$ against the adversary. It follows that $K^*$ is $\epsilon(n)$-private. $\Box$

We shall call this {\em privacy amplification} scheme {\em deterministic privacy amplification}
since it is deterministic and does not involve any communication between Alice and Bob.

\subsection{Putting Things Together}\label{final}

We are now ready to provide the final statement regarding the key authentication scheme described
in the previous sections. By key authentication process we loosely mean the procedures described in
Sections~\ref{method} and \ref{lost}. That is, it includes the deterministic privacy amplification
procedure run independently by Alice and Bob after the authentication process described in
Sect.~\ref{method} has resulted in a success: ${\tt result}={\tt result'}=1$.\\

\noindent{\bf Theorem 2.} {\em Let $K^*_A\in\{0,1\}^{n-s-m}$ and $K^*_B\in\{0,1\}^{n-s-m}$ be the
final secret keys generated after key authentication and deterministic privacy amplification as
described above upon initial $\epsilon(n)$--private $K^A,K^B\in\{0,1\}^{n}$. Suppose the MAC used
during key-authentication has impersonation probability at most $p_{{\tt im}}$ even given two
message-tag pairs authenticated with the same key. Then, against any ($\ell-1$)--bounded adversary we
have, \[ \Pr{({\tt result}= {\tt result'}=\delta_{K^*_A,K^*_B})} \geq
 (1-\epsilon(n))(1-2^{-m})(1-p_{{\tt im}})^{2\ell-2}, \] and $K^*_A$ and $K^*_B$ is $2^{-m} + 2\ell
 p_{{\tt im}} +2\epsilon(n)$--private. If in addition the adversary is ($\ell-2$)--bounded then the
 final secret key remains private the same way against honest-but-curious paths.}\\

\noindent{\em Proof.} The only thing that does not directly follows from Lemma 2 and 3 is the
statement about the privacy of $K^*_A$ and $K^*_B$. Privacy only makes sense when ${\tt
result}={\tt result'}=\delta_{K^*_A,K^*_B}=1$. When this applies however the final secret-key
$K^*=K^*_A=K^*_B$ is $\epsilon(n)$--private as it was shown in Lemma 3. The result follows
immediately. $\Box$

\subsection{What MAC to Use?}\label{mac}

Any authentication scheme with small enough impersonation probability $p_{{\tt im}}$ can be used by
Alice when she sends $M_A$. The authentication schemes used in {\sc Secoqc} follow~\cite{kraw, shoup}.
These authentication schemes can also be used for
key-authentication. However, the impersonation probability $p_{{\tt im}}$ should hold even given
two  message-tag pairs generated using the same key.

This can be achieved the obvious way by setting $\kappa_A=(\kappa'_A,\kappa''_A)=(K^A)_{1\ldots
2s}$ and $\kappa_B=(\kappa'_B,\kappa''_B)=(K^B)_{1\ldots 2s}$ in the key authentication process.
Alice authenticates message $M_A$ with sub-key $\kappa'_A$ which Bob verifies with sub-key
$\kappa'_B$. Bob's message $M_B$ is authenticated with sub-key $\kappa''_B$ while Alice verifies
with sub-key $\kappa''_A$. Clearly, if the $MAC$ scheme has impersonation probability at most
$p_{{\tt im}}$ given one message-tag pair then this way of authenticating as impersonation
probability at most $2p_{{\tt im}}$ against two message-tag pairs generated with the same key.
There are many other ways of building MACs suitable for our application\cite{Weg_Car}. The one
mentioned above is probably the simplest but certainly not the best one in terms of key size.


\section{Conclusions}
\label{sec:perspectives}

In this paper we have reviewed the concept of a QKD network and have discussed different models of
QKD networks. We have in particular focused on trusted repeater networks and have studied the case
when part of the nodes are not to be trusted and could be arbitrarily malicious. We have shown how
to ensure that Alice and Bob share identical and private keys after key generation  over the
network. We suppose that Alice and Bob do not share key material to start with. They only share
keys with their direct neighbours. However, we suppose that classical messages through honest paths
are eventually delivered to their intended recipient (assumption~(\ref{assumption})).

We conclude that secret keys can be generated through $\ell$ disjoint paths in a private and
authentic way against ($\ell-1$)--bounded adversaries and against ($\ell-2$)--bounded adversaries with
honest-but-curious paths.

It should be noted that assumption~(\ref{assumption}) can be relaxed further without undesirable
consequences for the security of the key authentication process. It suffices for only one honest
path to eventually deliver classical information to the  intended receiver. This does not modify by
any means neither the protocol nor its security analysis. Indeed, an honest path will always
allow parties to agree upon the authenticity of the secret key. Only one properly authenticated
message from Alice to Bob and one from Bob to Alice is sufficient to assess the equality of both
keys. Otherwise, if the keys are different then both parties will anyway conclude that keys do not
match.


\begin{thebibliography}{99}

\bibitem{gisin:rmp02}
N. Gisin, G. Ribordy, W. Tittel and H. Zbinden, ``Quantum Cryptography'', {\em Rev. Mod. Phys.}
{\bf 74}, 145 (2002).

\bibitem{dusek:pino06}
M. Du\^sek, N. L\"utkenhaus and M. Hendrych, ``Quantum Cryptography'' in {\em Progress in Optics},
vol. {\bf 49}, 381 (E. Wolf, Ed., Elsevier, 2006).

\bibitem{scarani:rmp08}
V. Scarani, H. Bechmann-Pasquinucci, N. J. Cerf, M. Dusek, N. L\"utkenhaus and M. Peev, ``A
Framework for Practical Quantum Cryptography'', {\em Rev. Mod. Phys.}, to be published (2009); eprint quant-ph/0802.4155 (2008).

\bibitem{Menezes}
A. Menezes, P. van Oorschot and S. Vanstone, {\it Handbook of Applied Cryptography} (CRC Press, Boca
Raton, 1997).

\bibitem{Shan1}
C. E. Shannon, ``A Mathematical Theory of Communication'', {\em Bell System Technical Journal}  {\bf 27},
379 and 623 (1948).

\bibitem{Shan2}
C. E. Shannon, ``Communication Theory of Secrecy Systems'', {\em Bell System Technical Journal}  {\bf 28},
656 (1949).

\bibitem{Wyner}
A. D. Wyner, ``The Wire-tap Channel'', {\em  Bell System Technical Journal}  {\bf 54}, 1355 (1975).

\bibitem{wolf:phd99}
S. Wolf, ``Information-theoretically and Computationally Secure Key Agreement in
Cryptography'', {\em ETH dissertation No. 13138}, (ETH Zurich, 1999).

\bibitem{teleport}
C. Bennett, G. Brassard, C. Crepeau, R. Jozsa, A. Peres, and W. Wootters,  ``Teleporting an Unknown
Quantum State via Dual Classical and Einstein-Podolsky-Rosen Channels'', {\em Phys. Rev. Lett.} {\bf 70}, 1895 (1993).

\bibitem{christandl:llncs07}
M. Christandl, A. Ekert, M. Horodecki, P. Horodecki, J. Oppenheim and R. Renner, ``Unifying
Classical and Quantum Key Distillation'', in {\em Proceedings of the 4th Theory of Cryptography
Conference}, Lecture Notes in Computer Science {\bf 4392}, 456 (S.P. Vadhan, Ed., Springer-Verlag, 2007).


\bibitem{secoqc}
http://www.secoqc.net

\bibitem{idsqmagiq}
http://www.idquantique.com, http://www.smartquantum.com, http://www.magiqtech.com


\bibitem{mayers:jcm01}
D. Mayers, ``Unconditional Security in Quantum Cryptography'', in {\em J. Assoc. Comput. Math.}
{\bf 48}, 351 (2001).

\bibitem{koashi:qp051}
M. Koashi, ``Simple Security Proof of Quantum Key Distribution via Uncertainty Principle'', eprint
quant-ph/0505108 (2005).

\bibitem{koashi:qp07}
M. Koashi, ``Complementarity, Distillable Secret Key, and Distillable Entanglement'', eprint
quant-ph/0704.3661 (2007).

\bibitem{gottesman:qic04}
D. Gottesman, H.-K. Lo, N. L\"utkenhaus and J. Preskill, ``Security of Quantum Key Distribution
with Imperfect devices'', {\em Quant. Inf. Comput.} {\bf 4}, 325 (2004).

\bibitem{lo:science283}
H.-K. Lo, H.F. Chau, ``Unconditional Security of Quantum Key Distribution over Arbitrarily Long Distances'',
{\bf 283}, 2050 (1999).


\bibitem{shor:prl00}
P. W. Shor and J. Preskill, ``Simple Proof of Security of the BB84 Quantum Key Distribution
Protocol'', {\em Phys. Rev. Lett.} {\bf 85}, 441 (2000).


\bibitem{renner:phd05}
R. Renner, ``Security of Quantum Key Distribution'', {\em Int. J. Quant. Inf. (IJQI)} {\bf 6}, 1 (2008).

\bibitem{kraus:prl05}
B. Kraus. N. Gisin and R. Renner, ``Lower and Upper Bounds on the Secret-Key Rate for Quantum Key
Distribution Protocols Using One-Way Classical Communication'', {\em Phys. Rev. Lett.} {\bf 95},
080501 (2005).

\bibitem{renner:pra05}
R. Renner, N. Gisin and B. Kraus, ``Information-theoretic Security Proof for
Quantum-Key-Distribution Protocols'', {\em Phys. Rev. A} {\bf 72}, 012332 (2005).

\bibitem{renner:llncs05}
R. Renner and R. K\"onig, ``Universally Composable Privacy Amplification Against Quantum
Adversaries'', in {\em Theory of Cryptography: Second Theory of Cryptography Conference, TCC 2005},
Lecture Notes in Computer Science {\bf 3378}, 407 (J. Kilian, Ed., Springer-Verlag, 2005).

\bibitem{konig:prl07} R. K\"onig, R. Renner, A. Bariska and U. Maurer, ``Small Accessible Quantum
Information Does Not Imply Security'', {\em Phys. Rev. Lett.} {\bf 98}, 140502 (2007).

\bibitem{Weg_Car} M. N. Wegman and J. L. Carter, ``New Hash Functions and Their Use in Authentication and Set Equality'', {\em J. Comp. Sys. Sci.} {\bf 22}, 265 (1981).

\bibitem{Car_Weg} J. L. Carter and M. N. Wegman, ``Universal Classes of Hash Functions'', {\em J. Comp. Sys. Sci.} {\bf 18},
143 (1979).

\bibitem{diffie:ieee76}
W. Diffie and M. E. Hellman, ``New Directions in Cryptography'', {\em IEEE Transactions on
Information Theory} {\bf 22}, 644 (1976).

\bibitem{alleaume:white07}
R. All\'eaume et al, ``SECOQC White Paper on Quantum Key Distribution and Cryptography'', eprint
quant-ph/0701168 (2007).

\bibitem{dianati:scn08}
M. Dianati, R. All\'eaume, M. Gagnaire, X. Shen, ``Architecture and Protocols of the Future
European Quantum Key Distribution Network'', {\em Security and Communication Networks} {\bf 1}, 57
(2008).


\bibitem{elliott:njp02}
C. Elliott, ``Building the Quantum Network'', {\em New J. Phys.} {\bf 4}, 46 (2002).

\bibitem{elliott:qp05}
C. Elliott, A. Kolvin, D. Pearson, O. Pikallo, J. Shlafer and H. Yeh, ``Current Status of the Darpa Quantum Network'', in {\em Quantum Information and Computation III}, Proc. SPIE {\bf 5815},  138 (E. Donkor, A. Pirich and H. Brandt, Eds., SPIE-The International  Society for Optical Engineering, 2005); eprint quant-ph/0503058 (2005).

\bibitem{Biham}
E. Biham, B. Huttner, and T. Mor, ``Quantum Cryptographic Network based on Quantum Memories'', Phys.
Rev. A, {\bf 54}, 2651 (1996).

\bibitem{briegel:prl98}
H.- J. Briegel, W. D\"ur, J. I. Cirac and P. Zoller, ``Quantum Repeaters: the Role of Imperfect
Local Operations in Quantum Communication'', {\em Phys. Rev. Lett.} {\bf 89}, 5932 (1998).

\bibitem{collins:jmo05}
D. Collins, N. Gisin and H. de Riedmatten, ``Quantum Relays for Long-distance Quantum
Cryptography'', {\em J. Mod. Opt.} {\bf 52}, 735 (2005).

\bibitem{Poppe:ijqi08} A. Poppe, M. Peev and O. Maurhart, ``Outline of the SECOQC Quantum-Key-Distribution Network in Vienna'', {\em  Int. J. Quant. Inf. (IJQI)} {\bf 6}, 209 (2008).



\bibitem{DDWY93} D. Dolev, C. Dwork, O. Waarts and M. Yung,
``Perfectly Secure Message Transmission'', {\em Journal of the ACM} {\bf 40}, 17 (1993).

\bibitem{DSECArch}
L. Salvail, ``Security Architecture for SECOQC: Secret Key Privacy and Authenticity over QKD
Networks'', {\em {\sc Secoqc} Deliverable} (2007), unpublished.

\bibitem{Sanders1}
T. R. Beals and B. C. Sanders, ``Distributed Authentication for Randomly Compromised Networks'', eprint
arXiv:0803.2917 (2008).

\bibitem{Sanders2}
T. R. Beals and B. C. Sanders, ``Distributed Relay Protocol for Probabilistic Information-Theoretic Security in a Randomly-Compromised Network''  in {\em Proceedings of International Conference on Information Theoretic Security (ICITS2008), Calgary, Alberta, 10 Aug 2008 - 13 Aug 2008''}, Lecture Notes in Computer Science {\bf 5155}, 29 (R. Safavi-Naini, Ed., Springer-Verlag, 2008); eprint arXiv:0803.2919.


\bibitem{DSEC}
L. Salvail and C. Schaffner, ``Rough Network Architecture for Quantum Communication Applied to
Basic Scenarios'', {\em {\sc Secoqc} Deliverable} (2004), unpublished.


\bibitem{DW02} I. Desmedt and Y. Wang, ``Perfectly Secure Message
Transmission Revisited'', in {\em Advances in Cryptology--Proceedings of Eurocrypt 2002}, Lecture
Notes in Computer Science {\bf 2332}, 502 (L. Knudsen Ed., Springer-Verlag, 2002).



\bibitem{DW06} I. Desmedt and Y. Wang, ``Perfectly Secure Message
Transmission Revisited'', unpublished (2006) - private communication.


\bibitem{kraw}
H. Krawczyk, ``LFSR-based Hashing and Authentication'', in {\em Proc. of CRYPTO'94}, Lecture Notes
in Computer Science {\bf 839}, 129 (Y. Desmedt, Ed., Springer-Verlag, 1994).

\bibitem{shoup}
V. Shoup, ``On Fast and Provably Secure Message Authentication based on Universal
Hashing'', in {\em Proc. Crypto'96}, Lecture Notes in Computer Science {\bf 1109}, 313  (N. Koblitz, Ed., Springer-Verlag, 1996).

\bibitem{alleaume:qp09} R. All\'eaume, F. Roueff, E. Diamanti and N. L\"utkenhaus, ``Topological optimization of
QKD networks'', eprint arXiv:0903.0839 (2009).


\end{thebibliography}
\end{document}